\documentclass[10pt,a4paper,dvips,twocolumn,showpacs]{revtex4}
\usepackage[english]{babel}
\usepackage{epsfig}
\usepackage{graphics, curves} 
\usepackage{amsmath, amsfonts, amssymb}
\usepackage{graphicx}
\usepackage{dcolumn}
\usepackage{dcolumn}
\begin{document}
\title{Capillary-Driven Instability of Immiscible Fluid Interfaces Flowing 
in Parallel in Porous Media}
\author{Thomas Ramstad}
\email[thomas@numericalrocks.com]{}
\affiliation{Department of Physics, Norwegian University of Science and
Technology, N--7491 Trondheim, Norway}
\affiliation{Numerical Rocks AS, Stiklestadveien 1, N--7041 Trondheim, Norway}

\author{Alex Hansen}
\email[Alex.Hansen@ntnu.no]{}
\affiliation{Department of Physics, Norwegian University of Science and
Technology, N--7491 Trondheim, Norway}
\affiliation{Numerical Rocks AS, Stiklestadveien 1, N--7041 Trondheim, Norway}

\date{\today}
\begin{abstract}
When immiscible wetting and non-wetting fluids move in parallel in a 
porous medium, an instability
may occur at sufficiently high capillary numbers so that interfaces between
the fluids initially held in place by the porous medium are mobilized.  A
boundary zone containing bubbles of both fluids evolve which has a well
defined thickness.  This zone moves at constant average speed towards the 
non-wetting fluid.  A diffusive current of bubbles of non-wetting fluid 
into the wetting fluid is set up.  
\end{abstract}
\maketitle

When a fluid displaces another one in a porous medium 
the interface separating the two fluids may become unstable.  
In the case of two-phase immiscible displacement, local
capillary barriers on pore-scale levels affect the behavior on larger scales, 
and it turns out that there is an extraordinary richness to
the ways instabilities occur and how the separating interface 
subsequently develops.  
Depending on several flow properties like viscosity ratio, 
wetting properties with respect to the
porous medium and how fast the displacement occurs, a wide range of
behaviors are found in both drainage and imbibition ranging 
from pure invasion percolation to viscous fingering \cite{d92,s95}. 
A huge effort has gone into classifying and understanding this rich
behavior, both from a fundamental scientific point of view, but also 
due to its importance in a number of very important fields ranging from
oil recovery, to spreading of pollutants in the ground water, to problems 
related to $\rm{CO}_2$ sequestering.

It is then surprising to discover that the related problem of immiscible
fluids flowing in parallel to the interface between them rather than 
normal to it in a porous medium, has received very little attention in 
comparison.  Such parallel flow is e.g.\ seen in connection with fully 
developed viscous fingers \cite{zy92} and in connection with flow in 
stratified reservoirs \cite{cda71,yl81,zl81,f88,fn88}.  When the flow rate is 
low so that capillary forces dominate at the interface, the 
parallel interface is
stable and each phase behaves as in a 
single-phase flow system. Nevertheless, it has been 
recognized that above a certain threshold in the flow-rate, but where capillary 
forces still dominate,
imbibition processes become important in the evolution of the 
interface and hence
the crossflow of the immiscible fluids.
However, at larger flow rates, where viscous forces dominate, shear-driven 
Kelvin-Helmholtz type instabilities are believed to occur 
\cite{b82,pb02,me04}.  Both theoretical and experimental work has been 
invested in studying the Kelvin-Helmholtz instability in vertical 
Hele-Shaw cells \cite{zy92,gr97,grrsw97}, as it provides a model for 
parallel flow in porous media in the viscous regime. 

It is the aim of this Letter to investigate the instability that occurs
at the interface between two immiscible fluids flowing in parallel in
a regime where capillary effects {\it cannot\/} be ignored.  This regime
has remained essentially untouched in the literature.  We find that      
above a threshold flow rate and with a viscosity ratio between the two
fluids favoring the formation of viscous fingers,
the interface becomes unstable, and a boundary zone appears containing 
intermixed bubbles of both fluids. 
This boundary zone has a well-defined width and 
moves at constant average speed towards the non-wetting fluid.  A diffusive current of 
bubbles of non-wetting fluid 
into the wetting fluid is set up, but the situation of
bubbles of wetting fluid entering the non-wetting fluid is absent. 

This instability may prove to be an important 
mechanism for mixing non-wetting fluid into wetting fluid.  A practical 
application may be CO$_2$ sequestering in porous rock formations.  The 
less wetting gas is blown into the porous medium which is already saturated 
by a more wetting fluid.  A mixing zone will then form at the boundary 
between the gas and the fluid where gas bubbles will be generated.
These bubbles are then transported into the wetting fluid where they
eventually are absorbed. 

We study this instability here using a two-dimensional network simulator first
developed by Aker {\it et al.\/} \cite{amhb98} with later extensions by 
Knudsen {\it et al.\/} \cite{kah02} and Ramstad and Hansen \cite{rh06}. 
The network forms a square lattice oriented at 45$^\circ$ with respect to the
overall flow direction.  Each link forms an hour-glass shaped tube. 
Disorder is introduced in the model by having the average tube radii $r$
be drawn from a flat distribution on the interval  
$r \in (0.1\ell, 0.4\ell)$, where $\ell$ is the link length. Capillary 
pressure in the links is caused by the presence of interfaces in them.

As the tubes are hour glass shaped,
the capillary pressure difference caused by a meniscus at
position $x$, the distance from one of the two nodes it is 
attached to, is given by $p_c \propto 1-\cos(2\pi x/\ell)$.

We assume cylindrical tubes so that the flow rate $q$ in a tube is 
given by the Hagen-Poiseulle relation from laminar flow
\begin{equation}
q = - 
\frac{\pi r^4}{8\ell\mu_{\rm eff}} \left(\Delta p - \sum p_c \right) \ ,
\end{equation}
where $\Delta p$ is the pressure difference between the nodes
connected by the tube.
The effective viscosity is the volume-weighted average of the
viscosities of the fluids contained in the tube. The sum runs over
the number of meniscii in the tube.  We accept up to 10 meniscii in any
given tube.  If this number is exceeded or the distance between two  
bubbles is too small, we merge the mensicii.  

The flow equations are solved by assuming flux conservation at each
node, i.e., invoking the Kirchhoff equations.  This is done by defining
a pressure $p$ at each node. We use the conjugate gradient method for this
\cite{bh88}.  After the node pressures have been determined, the positions
of the meniscii are integrated forwards by an adaptive time-step 
$\Delta t$ so that no single meniscus movement exceeds one tenth
of a tube-length $\ell$.  When meniscii reach the ends of a tube, 
they are moved into the other eligible tubes connected to that node.
For details, see \cite{kah02}.

The flow of the two fluids in the network is controlled by the ratio 
between capillary and viscous forces at the pore level and
quantified through the capillary number
\begin{equation}
Ca = \frac{\mu Q_{\rm tot}}{\gamma \Sigma} \ ,
\label{ca}
\end{equation}
where $\mu$ is the largest viscosity of the two immiscible fluids, 
$\Sigma$ is the cross-sectional area
of the network and $Q_{\rm tot}$ is the total flux through this area.

Besides the capillary number, the ratio between the viscosities of the
two fluids forms the second dimensionless number to control the flow, 
\begin{equation}
M = \frac{\mu_{\rm nw}}{\mu_{\rm w}} \ .
\end{equation}
We set $M=1$ in the following.  Hence, there is initially no pronounced 
shear in the flow patterns in the network.

We implement periodic boundary conditions in the average flow direction
\cite{kah02,rh06}.  This implies that the flow configurations sees no 
boundaries in the flow direction, and the fluid configurations may
develop over large times and distances.  There is no periodicity in the
direction normal to the average flow direction.  The boundaries
parallel to the average flow direction are in contact with a reservoir
of either wetting or non-wetting fluid. A constant pressure drop $\Delta P$
is set up across the network in the average flow direction, causing the
total flux $Q_{\rm tot}$. 

\begin{figure}[]
\includegraphics[width=0.45\columnwidth,clip]{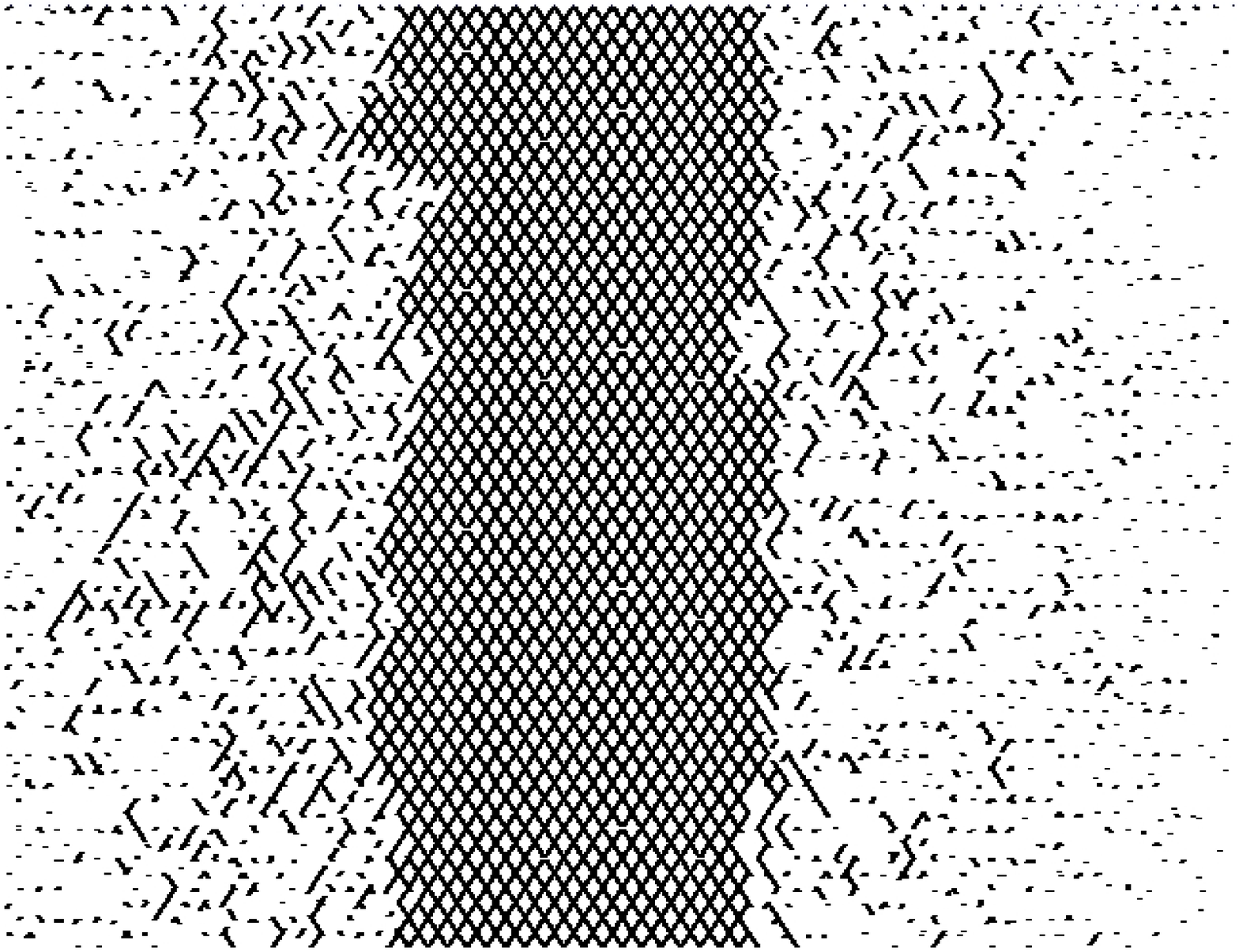} 
\hfill
\bigskip
\includegraphics[width=0.45\columnwidth,clip]{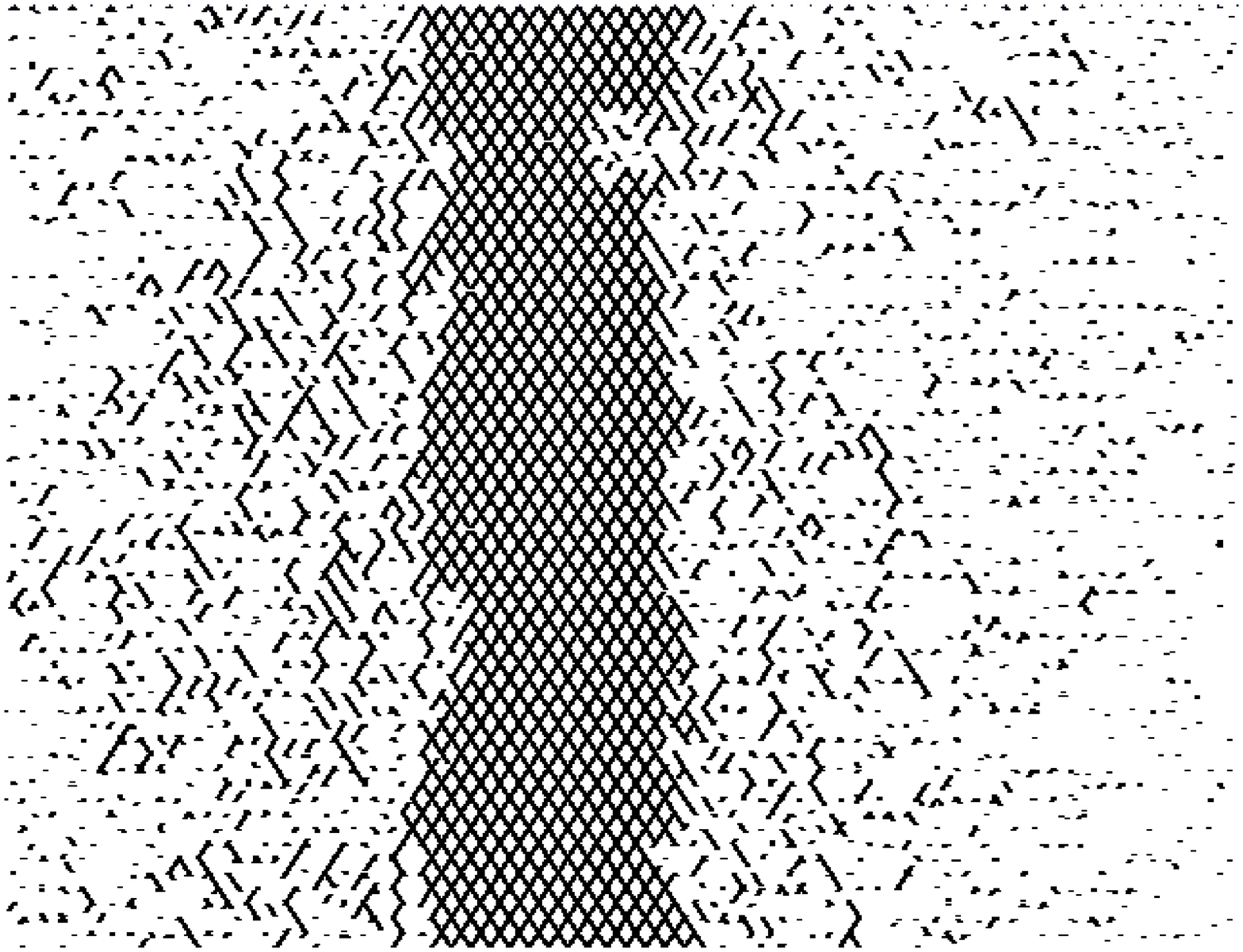}
\bigskip
\includegraphics[width=0.45\columnwidth,clip]{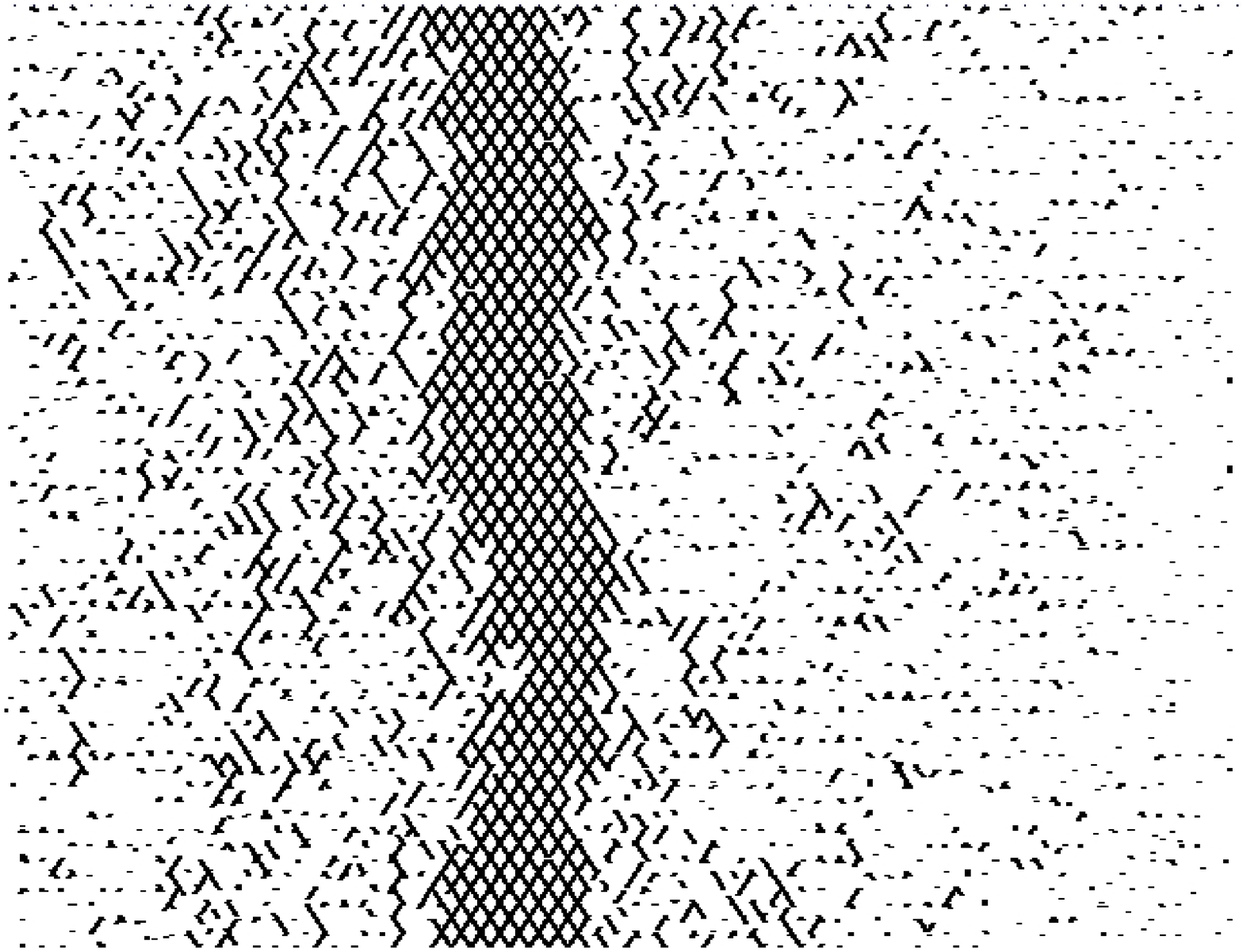}
\hfill
\includegraphics[width=0.45\columnwidth,clip]{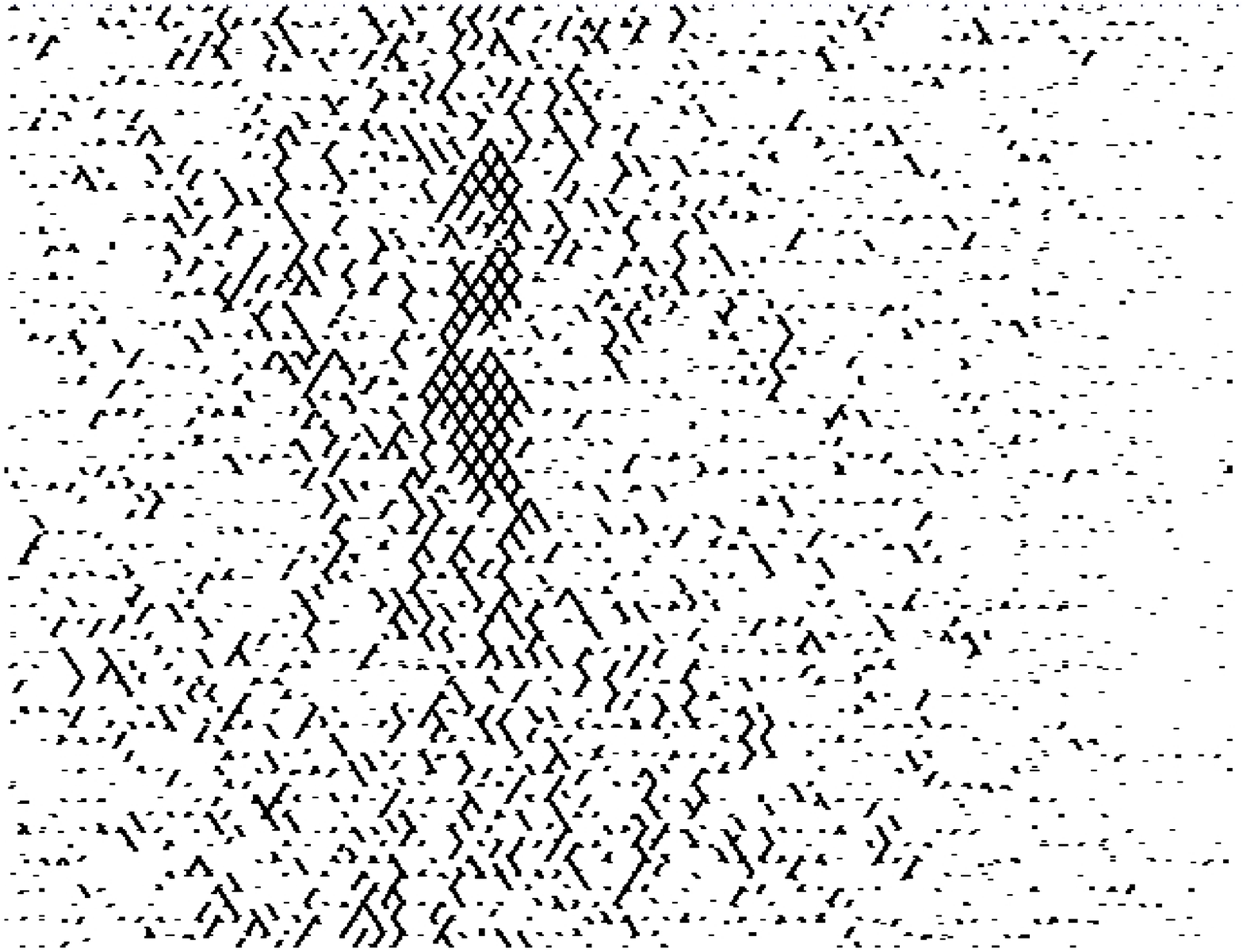}
\caption{\label{fig1} Different stages of the development
of an initially straight band of non-wetting fluid (black) inside
a region filled with wetting fluid (white).  The flow is from top to bottom
with periodic boundary conditions this direction. The boundaries are open
in the transverse direction.  The size of the network is 
$L_x \times L_y = 64 \times 64$.}
\end{figure}

The network is prepared with either a band of non-wetting fluid parallel to the 
average flow direction, surrounded by wetting fluid or {\it vice versa.\/}  
Hence, the saturation of non-wetting fluid, $S_{\rm nw}$ and 
wetting fluid, $S_{\rm w}$ is non-uniform. We show in Fig.\ \ref{fig1} 
the network initially prepared with a band of non-wetting fluid in the 
middle.  If the 
pressure drop $\Delta P$ is too small, the interfaces in the tubes forming the
boundaries between the two fluid types will be stabilized by the capillary
pressures and the boundaries are stable.  However, when the pressure difference
is above a minimal value so that the initial capillary number $Ca_{\rm init} > Ca_{\min}$,
the boundaries destabilize and the system evolves.

\begin{figure}[t!]
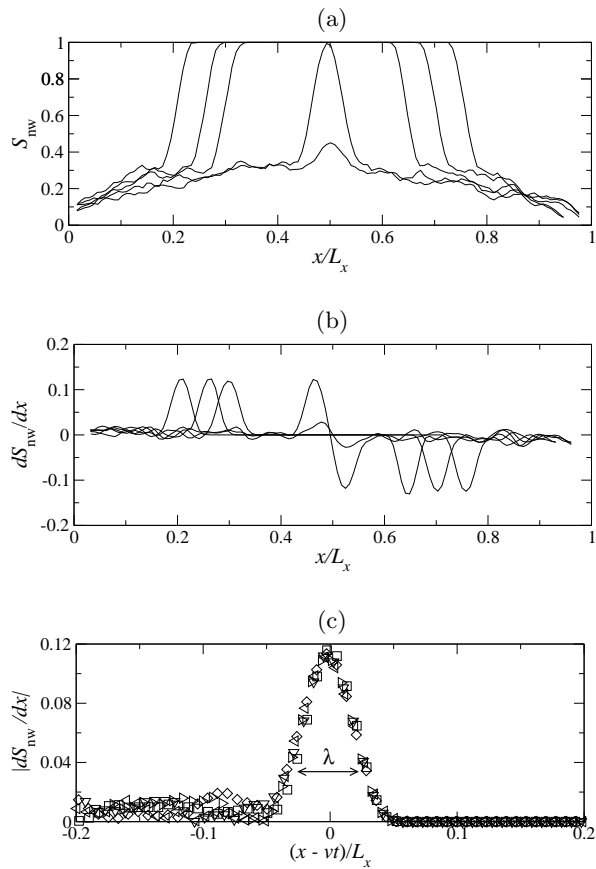

\begin{center}
\ \ \ \ \ \ \ (a) \\
\includegraphics[width=0.9\columnwidth,clip]{rh-fig2a.eps} \\
\ \\
\ \ \ \ \ \ \ (b) \\
\includegraphics[width=0.9\columnwidth,clip]{rh-fig2b.eps} \\
\ \\  
\ \ \ \ \ \ \ (c) \\
\includegraphics[width=0.9\columnwidth,clip]{rh-fig2c.eps}\\
\end{center}
\caption{\label{fig2} (Color online) 
(a) $S_{\rm nw}$ {\it vs.\/} $x$ for different $t$,  (b) $d S_{\rm nw}/dx$
{\it vs.\/} $x$ for different $t$, (c) $dS_{\rm nw}/dx$ {\it vs.\/}
$(x - vt)/L_x$ for different $t$ for a moving front with starting point $x_0 = 0$. 
All figures are 
for an $L_x \times L_y = 128 \times 32$ lattice
with open boundaries in the direction parallel to the overall flow and with
$Ca_{\rm init} = 0.03$.}
\end{figure}

Early in the evolution of the system,
fingers of non-wetting fluid form when the viscosity ratio between the two
fluids allow this. This is a signature of unstable non-wetting
front propagation in the viscous regime.  The fingers are bent in 
the direction of the average flow.  Due to the flow typically being at 
an angle compared to the fingers, they are susceptible to break up. The broken
off fingers form bubbles that migrate into the wetting fluid, and consequently 
the wetting 
fluid also migrates into the non-wetting fluid.  This is due to the appearance
of an effective pressure gradient $\Delta P_{\bot}$ normal to the average 
flow direction across the boundary region between the two fluids.  This
gradient is in turn due to the appearance of a gradient in the effective
permeability.  The effective pressure gradient $\Delta P_{\bot}$ leads to 
imbibition of the wetting fluid into the non-wetting region.  This process 
creates a compact front and a saturation profile moving in the direction
normal to the average flow direction, resembling that of 
Buckley-Leverett flow \cite{s95}.  

There is a length scale $\lambda$ associated with the saturation profile.
We define it through the width of the bell-shaped non-wetting saturation
gradient as shown in Fig.\ \ref{fig2} based on an average over five 
samples.  From the motion of the two maxima of $dS_{\rm nw}/dx$ along the 
$x$-axis, which is the direction normal to the average flow direction, we
determine the mean velocity of the non-wetting saturation profile.  The data
collapse shown in Fig.\ \ref{fig2}c,
where $dS_{\rm nw}/dx$ is plotted against $(x- vt)/L_x$, shows that the mean
velocity $v$ of the profile is constant and the shape of the profile is 
also constant. The length scale $\lambda$ is for this system 
$\lambda/L_x\approx 0.04$.  Hence, unlike the Kelvin-Helmholz shear 
instability, the interfacial instability between two different fluids in a 
porous medium proceeds through the creation of a well-defined saturation 
profile characterized by a length scale $\lambda$ and an average speed $v$ which
remains constant.

The shape of the non-wetting saturation profile corresponds to a there being
a boundary region where bubbles are created.  This boundary region 
moves into the non-wetting zone.  There are no bubbles migrating into 
this zone ahead of the moving boundary region.  On the other side, 
there is a diffusion current of bubbles of non-wetting fluid into the 
wetting zone.  The diffusing bubbles stem from the non-wetting fingers
that break off due to the average flow being at an angle with respect to
the fingers.  

When the two approaching boundary regions eventually meet, the middle
non-wetting band is destroyed as shown in the last picture of the sequence
shown in Fig.\ \ref{fig1}.
 
\begin{figure}[b!]
\begin{center}
\includegraphics[width=0.9\columnwidth,clip]{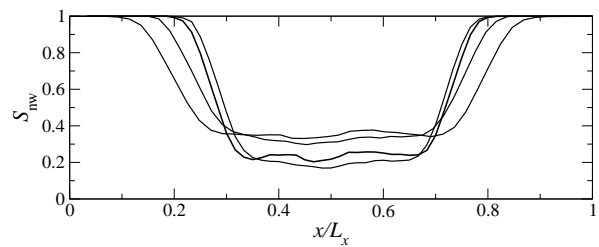} \\
\end{center}
\caption{\label{fig3} Non-wetting saturation profiles 
for the initial configuration of a band of wetting fluid
surrounded by non-wetting fluid. The profile 
stabilizes at different levels for different $\Delta P$.
For higher $\Delta P$ the saturation 
$S_{\rm nw}$ in the boundary region stabilizes at a higher level and
the wetting front advances further.}
\end{figure}

We now reverse the initial configuration so that 
a band of wetting fluid is surrounded by non-wetting fluid. 
We show in Fig.\ \ref{fig3} the evolution of the non-wetting 
saturation profile as a function of time.  As before, 
boundary regions where bubbles form are created.  However,
after some initial time, they stabilize and do not move.
This is in sharp contrast to the previous situation where
the boundary regions move with constant mean velocity.  This is
caused by there being no bubble transport 
outside the boundary region and into the non-wetting region.
Inside the wetting band, there is diffusive bubble transport, 
but as the width of the band  is finite and it is surrounded 
by bubble-generating boundary regions on both sides, the net
diffusive current stabilizes at zero.  

\begin{figure}[]
\begin{center}
\includegraphics[width=0.9\columnwidth,clip]{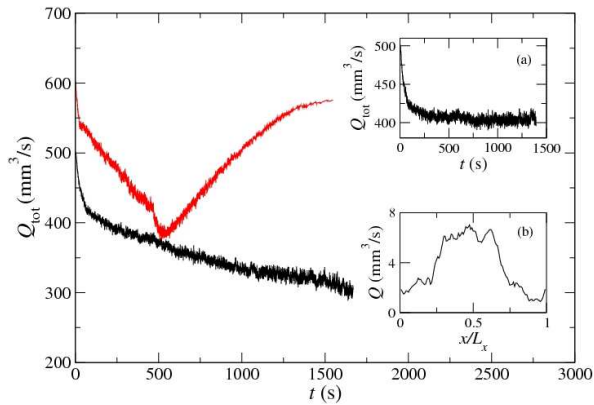} \\
\end{center}
\caption{\label{fig4} (Color online) $Q_{\rm tot}$ as a function 
of time for two 
different constant pressure differences $\Delta P$ when the middle
band is non-wetting.  Inset (a) shows the same for one constant
pressure difference when the middle band is wetting.  Inset (b)
shows the flux profile normal to the flow profile for an initial
non-wetting band in the middle.}     
\end{figure}

We consider in the following the evolution of the total flow rate 
$Q_{\rm tot}$ as the system evolves for both configurations we have studied.
We consider first the case of a non-wetting fluid band surrounded by wetting
fluid. As the flow is sustained by a constant pressure drop across the
network in the average flow direction, $\Delta P$, the total flow rate
$Q_{\rm tot}$ will at all times be proportional to the permeability
of the network. We show in Fig.\ \ref{fig4} the evolution of the 
total flow rate as a function of time for networks initially prepared with
a non-wetting band in the middle and with a wetting band in the middle. 
We analyze first the case when the networks starts with a non-wetting band
in the middle.  The evolution of $Q_{\rm tot}$ for two different pressure
drops are shown in Fig.\ \ref{fig4}.  We see that for both pressure drops,
$Q_{\rm tot}$ decreases linearly in time after an initial transient.  
In the case
of the larger pressure drop, the red curve, the flow rate starts increasing
again reaching essentially the flow rate it had initially.  This behavior
can be understood as follows.  After the initial transient and before the
rapid increase of $Q_{\rm tot}$ in the high-$\Delta P$ case, the system
consists of three zones: (1) a non-wetting zone characterized by an 
effective local permeability $k_{\rm nw}$, (2) a boundary zone characterized
by a local permeability $k_{\lambda}$ and (3) a zone where non-wetting fluid
forms diffusing bubbles in the wetting fluid.  The local permeability here
is $k_{\rm mix}$.  If the width of the non-wetting zone is $\ell_{\rm nw}$,
of the boundary zone is $\lambda$ and of the mixed zone is
$\ell_{\rm mix}$, then the total permeability of the network is given by
\begin{equation}
\label{sumk} 
k_{\rm eff}  = k_{\rm mix} \frac{2\ell_{\rm mix}}{L_x}
              +k_{\lambda} \frac{2\lambda       }{L_x}
              +k_{\rm nw}  \frac{ \ell_{\rm nw} }{L_x}\;.  
\end{equation}
As the boundary regions moves with constant average speed $v$, we have that
$\ell_{\rm mix} = \ell_{\rm mix, 0} + vt$ and since $L_x = 2\ell_{\rm mix}
+2\lambda + \ell_{\rm mw}$, it follows that $\ell_{\rm nw} = L_x - 
2\ell_{\rm mix,0}-2\lambda -2vt$.  Inserting these two equations
in Eq.\ (\ref{sumk}) gives 
\begin{equation}
\label{kt}
k_{\rm eff} = k_{\rm eff,0} -\frac{2vt}{L_x}\ 
\left[k_{\rm nw} - k_{\rm mix}\right]\;.
\end{equation}
Since the viscositities of the two fluids are equal, $Q_{\rm tot} \propto
k_{\rm eff}$.  As $k_{\rm nw}$ is larger than $k_{\rm mix}$, the total flow 
rate $Q_{\rm tot}$ falls off with time.   

Fig.\ \ref{fig4} shows that at a larger pressure difference, the total
flow rate starts increasing again after the linear regime we have just 
described.  This is due to the non-wetting band in the middle having been 
depleted and the non-wetting bubbles are diffusing out of the network.  
Then the network is being depleted of non-wetting fluid and hence
interfaces which lowers the effective permeability.
The opposite situation, a middle wetting band, is shown in inset (a) in
Fig.\ \ref{fig4}.  We see that the total flow rate saturates, indicating that
the system enters a steady state as already discussed in connection with 
Fig.\ \ref{fig3}.

We have in our numerical experiments kept the pressure difference
across the network constant.  If we rather had kept the total flow 
rate $Q_{\rm tot}$ constant, the instability that sets in when $Ca >
Ca_{\min}$ will be much more violent.  This is so since the pressure
drop $\Delta P$ will increase to keep $Q_{\rm tot}$, leading in turn
to an acceleration of the boundary region.  

In this Letter we have investigated the stability where 
two different fluid flow parallel to each other.
We find that under constant pressure conditions, for a sufficiently
high capillary number a boundary region develops with a well-defined
width when the viscosity ratio between the two fluids favor 
the formation of viscous fingers.  
This region, which essentially is foam, moves at a constant
average speed into the non-wetting region.  On the wetting side, a diffusive
current of non-wetting bubbles away from the boundary zone develops.    
It would be of great interest to see this instability reproduced 
in the laboratory, e.g.\ in two-dimensional glass-bead filled Hele-Shaw
cells.

We acknowledge very interesting and useful discussions with 
E.\ G.\  Flekk{\o}y, H.\ A.\ Knudsen, K.\ J.\ M{\aa}l{\o}y and P.-E.\ {\O}ren. 
This work has been 
supported by Norwegian Research Council Grant No. 154535/432. A.\ H.\ thanks
Numerical Rocks AS for their hospitality.



\end{document}